**Spatial Cross-Recurrence Quantification Analysis for Multi-Platform Contact Tracing and Epidemiology Research**

K. Jakob Patten

College of Health Solutions, Arizona State University

**Author Note**

K. Jakob Patten  https://orcid.org/0000-0002-7931-5413

I have no conflict of interest to disclose.

Correspondence: Coor Hall 976 S. Forest Mall, Tempe, AZ 85287. Email: kjp@asu.edu

**Abstract**

Contact tracing is an essential tool in slowing and containing outbreaks of contagious diseases. Current contact tracing methods range from interviews with public health personnel to Bluetooth pings from smartphones. While all methods offer various benefits, it is difficult for the different methods to integrate with one another. Additionally, for contact tracing mobile applications, data privacy is a concern to many as GPS data from users is saved to either a central server or the user's device. The current paper describes a method called spatial cross-recurrence quantification analysis (SpaRQ) that can combine and analyze contact tracing data – regardless of how it has been obtained – and generate a risk profile for the user without storing GPS data. Furthermore, the plots from SpaRQ can be used to investigate the nature of the infectious agent, such as how long it can remain viable in air or on surfaces after an infected person has passed, the chance of infection based on exposure time, and what type of exposure is maximally infective.

**Spatial Cross-Recurrence Quantification Analysis for Multi-Platform Contact Tracing and Epidemiology Research**

A form of spatial analysis and contact tracing was used as early as the 1854 cholera outbreak in London. Anaesthetician and epidemiologist John Snow plotted incidences of the disease on a map of the Soho district.[1,2] This exercise led him to the hypothesis that cholera might be a water-borne illness and, furthermore, that the Soho outbreak might have come from a single water pump. This was complicated, however, by the existence of several cholera cases that were closer to other community water pumps. Snow traced how these outlying families obtained their water and found that, even though the households were closer to other pumps, the families prefered the taste of the Broad Street pump. Not only did Snow's efforts pinpoint the source of the outbreak, they also lead to the champion of the germ theory of disease over the miasma theory, and the development of epidemiological methods.

Contact tracing is an essential tool in epidemiology because it allows public health personnel to track and contain potential disease vectors and arrest the spread.[3-6] Outbreaks of diseases such as ebola and HIV have been contained and, in the case of smallpox, eradicated with the aid of contact tracing. Disease testing, public education, and – if available – vaccinations are also important factors in controlling the spread of a disease but contact tracing helps reduce the infected population to more managable levels. Health experts note that, especially for respiratory diseases like the current COVID-19 health crisis, early identification and isolation of cases is integral to overcoming the disease.[7,8]

An effective contact tracing tool must address three large problems facing the technology. First, while contact tracing has traditionally been conducted through one-on-one interviews between an infected person and a public health official, more technological means have been developed in recent years, especially in the wake of COVID-19.[9,10] Depending on the method used to code for spatial position – proprietary GPS data in a mobile smartphone application or addresses in an in-person interview, for instance – it can be difficult to combine these tracing methods into a more comprehensive, community-wide analysis of disease vectors. Effective contact tracing must integrate with multiple data gathering platforms. Second, successful contact tracing must prioritize patient privacy and security of user data. Technologists have warned that central databases of location data are vulnerable to breaches, while contact data stored on individual devices – such as the Bluetooth ping design – can present security weakpoints for those devices and other data contained therein.[9,11] Finally, to help prevent future outbreaks instead of simply managing a current one, contact tracing tools must – as John Snow's map supported the germ theory of disease – allow for new knowledge to be gleaned about the disease. Spatial cross-recurrence quantification analysis (SpaRQ) is amenable to multiple data types, stores location data for patients with positive results but does not need to store user data, and can provide insight into transmission vectors.

**Recurrence Quantification Analysis**

Traditional recurrence quantification analysis (RQA) is typically used to assess the similarity of a signal to itself.[12] The signal, a unidimensional vector such as distance from a central point or notes in a song, is plotted on both the *x*- and *y*-axes. Each individual moment in time is compared against all others. Where the value is the same, the intersecting box is valued as 1; where the value differs, the intersecting box is valued as 0. This creates a binary matrix where 1s indicate similarity and 0s indicate difference. Figure 1 illustrates the self-similarity in the traditional songs *Happy Birthday* and *Frère Jacques*. In both, the time series is made up of notes from the melody of the song, with each note treated as a single time point regardless of actual note duration or rests/silences in the sheet music.

In RQA, the solid, central diagonal line is meaningless because this is where the time series is plotted against itself in the same time window. In this way, an RQA plot is like a correlation matrix in that the main diagonal is an identity line. Solid diagonal lines off center, however, indicate places where the musical phrases are being repeated. This is especially evident in the repetitive phrases in *Frère Jacques.* One common metric used to assess similarity is the recurrence rate, the percentage of points in the matrix where a value recurs. Determinism is a related metric that returns the percentage of points that form a diagonal line. The ratio between determinism and recurrence can also be used to quantify spurious synchronization and prolonged synchronization. Finally, the aptly named maxline, which is simply the longest continuous non-center diagonal line, provides a measure of how precisely a time series replicates itself.

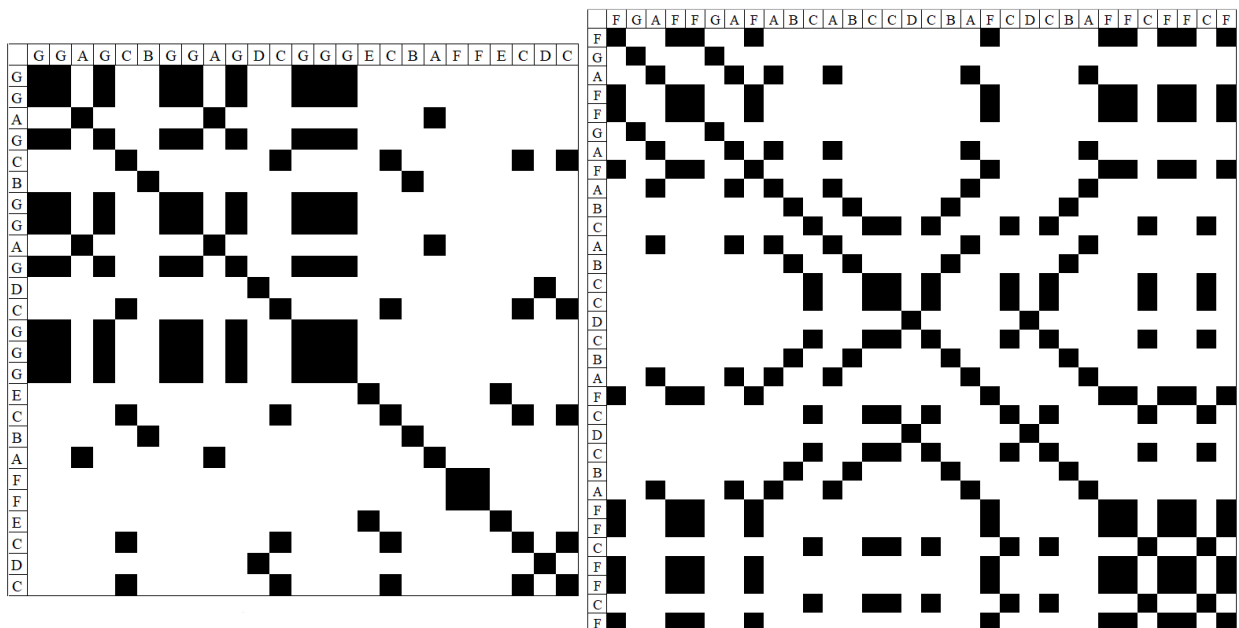

*Figure 1.* Recurrence plots for *Happy Birthday* (left) and *Frère Jacques* (right). Note that the upper left corner of *Frère Jacques* reveals the repetition of the first four notes while the longer diagonal lines below the midpoint highlight the repetition of the longer third verse.

Cross-recurrence quantification analysis (CRQA) is similar to RQA except two different signals are plotted against one another, rather than one signal against itself.[12,13] Figure 2 shows *Happy Birthday* plotted against *Frère Jacques*. In CRQA, the central diagonal becomes meaningful; it reveals when the two signals are in perfect synchronization with no time lag. In terms of a song, it would mean a portion of the melody is constructed with identical notes occurring at an identical point in the song. Off-center diagonal lines, as in RQA, carry the same meaning as lines on the central diagonal, but with a time lag. The same quantification metrics used in RQA can be used in CRQA.

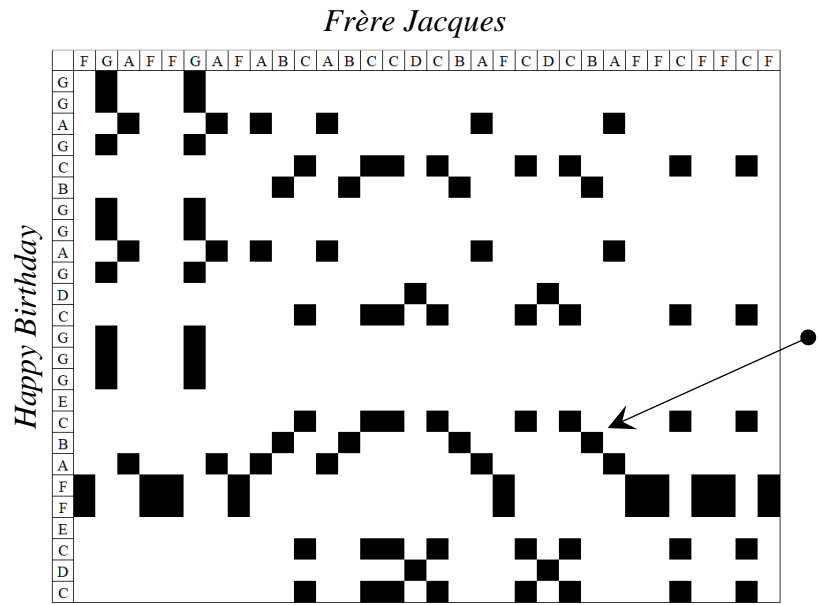

*Figure 2.* Cross-recurrence plot of *Happy Birthday* and *Frère Jacques.* The diagonal lines in the lower right quadrant reveal musical similarity between the, "…dear Ja-acques. Hap-py…," and, "…bells are ring-ing. Morn-," portions of the songs.

## Spatial Cross-Recurrence Quantification Analysis (SpaRQ)

While RQA and CRQA are often used in dynamical systems research to investigate the nonlinear relationships within a signal or between two signals, the same quantification methods can be used to analyze spatial proximity between two mobile agents. The current paper refers to this process as SpaRQ. Figure 3 shows the trajectories for two mobile agents, one infected and one healthy. The paths of these two agents cross multiple times; they are in the same place at the same time, they are in same place at different times, they are in the same place on one spatial axis (such as longitude) but not the other, and they are in different places at different times.

SpaRQ will turn these movement vectors into more intelligible plots and quantified proximity values that can be used to construct risk metrics and uncover transmission behavior of the disease under investigation.

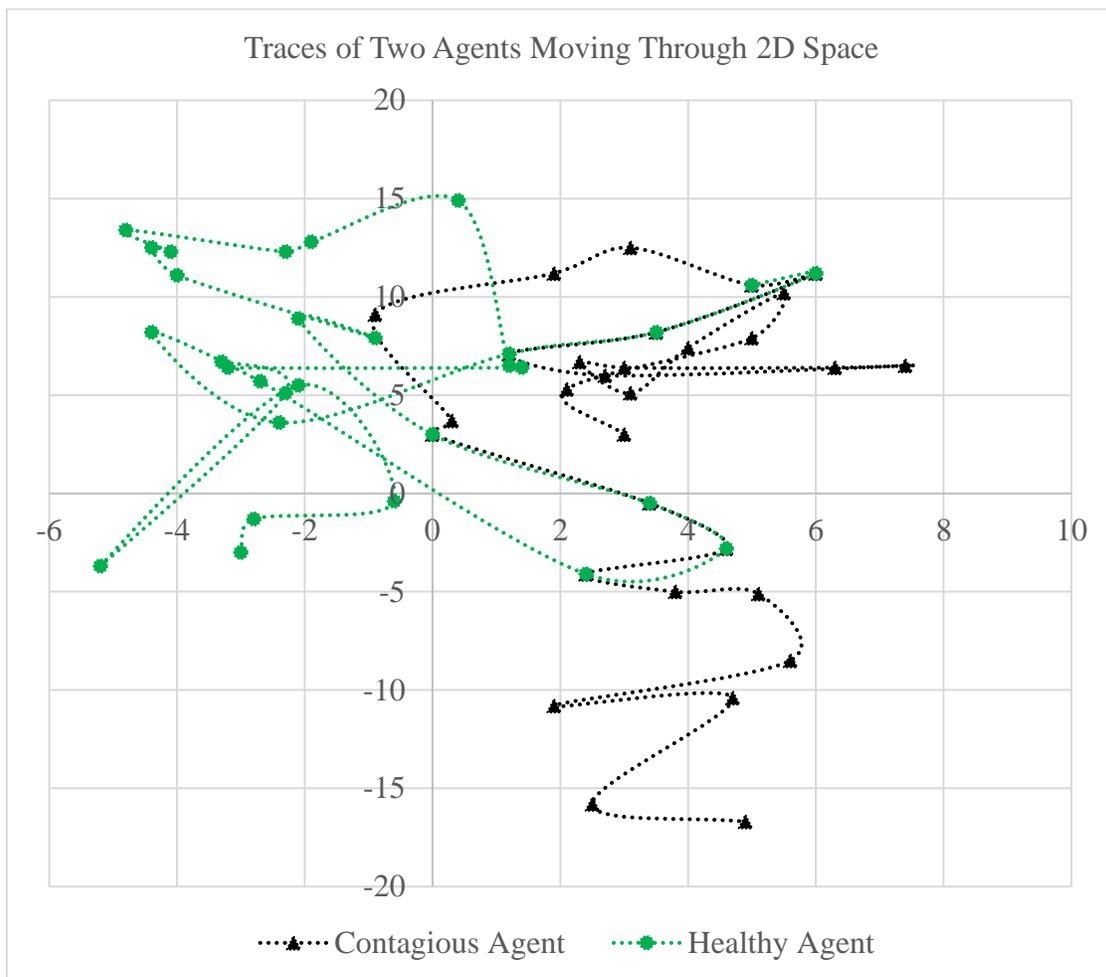

*Figure 3.* Traces of two agents moving through two-dimensional space. Agents coincide with one another in space with no time lag, with time lags, and in only one spatial dimension or the other.

Unlike the musical notes used in the previous examples, position data is time-dependent.

Continuous position data is averaged over a window, thus keeping both the infected and healthy agent on the same timetable. The size of this window is not a constant, as diseases are differentially infective and the amount of time needed to constitute a risk of infection varies greatly by condition. The CDC suggests a rough average of 15 minutes of close, non-contact exposure constitutes a risk for COVID-19.[14] Because position data is averaged throughout the time window, it is possible that two agents could be in close proximity to one another for the requisite time, but out of sync with the temporal slicing. Reducing the temporal window in accordance with the Nyquist-Shannon theorem to half or less the critical exposure time will offer more precise results.[15]

Depending on how the information is obtained, real-world position data may be continuous, as well. If the data come from an interview, the spatial precision is likely to be no more than a street address. If obtained from a user's Google Maps account, the data are likely to be much finer grained. To reduce false positive for contact, position data should be rounded to values corresponding with airborne transmission distances where possible. These distances are not yet determined for COVID-19, and can be anywhere between 2 and 10 meters.[14,16]

In addition to being continuous, geographical location is also a two- (and sometimes three-) dimensional data point. As such, standard RQA and CRQA cannot be used. Instead, the dimensions must be temporarily split into different data sets. With SpaRQ, the $x$- and $y$-coordinates (latitude and longitude, respectively, assuming this dataset represents GPS coordinates) for both agents are separated and grouped by type (i.e., $x_{infected}$ paired with $x_{healthy}$). The resulting groups are then plotted against one another as in CRQA. This process can be seen in Figure 4.

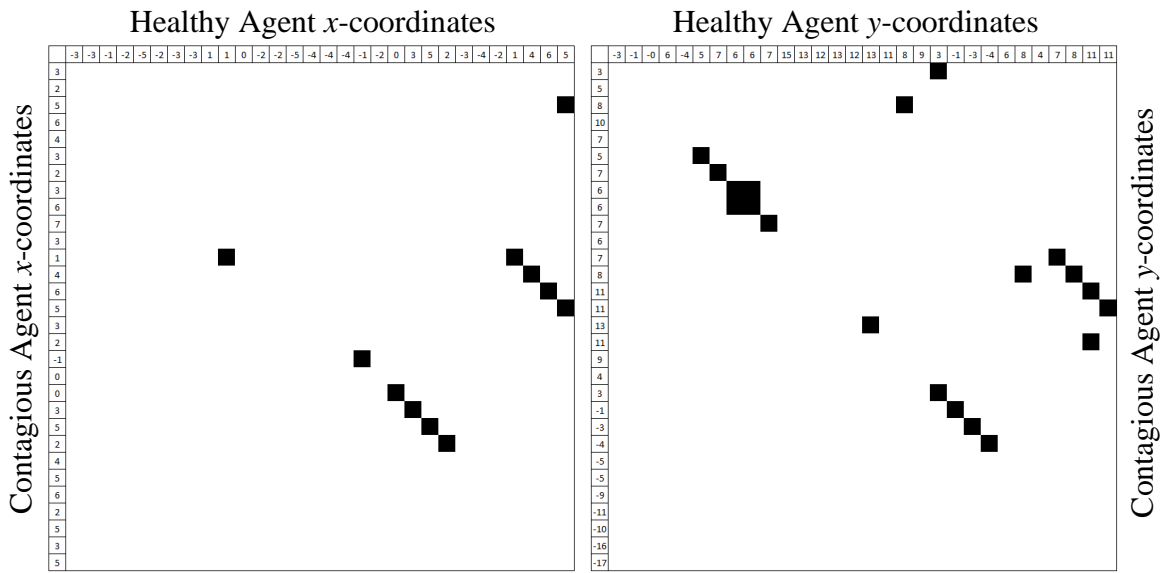

*Figure 4.* Cross-recurrence plots for *x*- and *y*-dimensions of spatial data from Figure 3. Note that there are similarities and differences in the resulting plots. The long central diagonal in the upper left corner of the *y*-coordinate plot is not replicated in the *x*-coordinate plot. This means the contagious and healthy agent were in the same *y*-plane, but not proximal on the *x*-plane.

The resulting *x*- and *y*-coordinate plots show the spatial synchronization between the contagious and healthy agent in those specific dimensional planes. For instance, there can be a large amount of overlap in the *x*-dimension and none in the *y*. This does not imply the two agents were near one another. To create a single matrix that implies proximity, the resulting *x*- and *y*-matrices must be combined by multiplying each square by its other-dimension counterpart. Because the matrices are constructed of only 0s and 1s, proximity in a single plane is nullified. The central diagonal of the combined plot is the proximity of both agents with no time lag. The upper triangular matrix is the proximity of both agents when the healthy agent lags the contagious agent. Both the central diagonal and the upper triangular matrix are important for assessing risk and analyzing the infectiousness of the traced disease. The lower triangular matrix, where the contagious agent lags the healthy, is of no use and can be deleted to save on computation and file size. The combination of the two matrices from Figure 4 can be seen in Figure 5. Finally, multiple proximity matrices computed for a single healthy agent against multiple contagious agents can be additively combined to produce a more complete picture of one person's risk of infection.

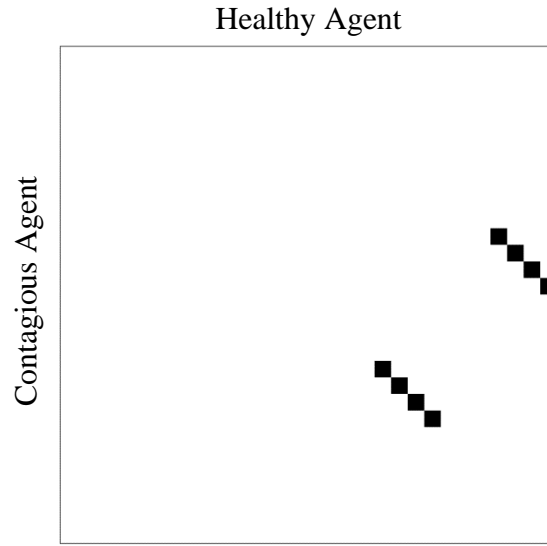

*Figure 5.* The result of combining *x*- and *y*-coordinate matrices from Figure 4. The central diagonal line indicates when the two agents were in the same place at the same time. The off-center line indicates when the healthy agent passed through space previously occupied by the contagious agent. The off-center line is 8 units away from the center. With a large dataset and knowledge of which healthy agents later became infected, the distance of lines from the center diagonal can be used to assess non-simultaneous infectivity of the traced disease.

The combined spatial matrix can now be quantified. Many traditional quantification methods of RQA and CRQA are useful in SpaRQ, albeit with slight modifications. Unlike traditional RQA or CRQA, the total number of discrete timepoints tested is not important. When looking at recurrence within a song, 12 notes recurring has a far different meaning for a song with 30 total notes than one with over 1000. In terms of contact tracing, the total exposure time is the most important value, not the percentage exposure time. As such, computing any metrics as a proportion may hinder the usefulness of the SpaRQ method. Likewise, as previously mentioned, all metrics used in SpaRQ are computed only for the upper triangular matrix, as the lower triangular matrix indicates where the infected agents lags the healthy. In the following equations, $N$ refers to the total number of points in the matrix, $i$ refers to the contagious agent's spatial series, $j$ refers to the healthy agent's spatial series, and $P(\ell)$ refers to a histogram of diagonal line lengths. Contact$_{Total}$, similar to recurrence rate in typical RQA and shown in (1), is a proportional

$$C_{Tot} = \frac{1}{.5N^2 + .5N} \sum_{i=1, j \geq i}^{N} \mathbf{R}(i, j) \qquad (1)$$

value that quantifies instances in which the healthy agent crossed paths, either simultaneously or at a time lag, with the contagious agent. Contact$_{TotalRaw}$, shown in (2), is the raw count of time

$$C_{TotRaw} = \sum_{i=1, j \geq i}^{N} \mathbf{R}(i, j) \qquad (2)$$

units in which the contagious and healthy agent crossed paths. Likewise, Contact$_{Sustained}$, similar to determinism in typical RQA and shown in (3), quantifies the amount of sustained direct or

$$C_{Sus} = \frac{\sum_{\ell = \ell_{min}}^{N} \ell P(\ell)}{\sum_{\ell = 1}^{N} \ell P(\ell)} \qquad (3)$$

lagged contact between the two agents. In (3), $\ell_{min}$ refers to the minimum length identified as a significant span of time. For instance, if 15 minutes is identified as the minimum time of interest and time was represented at a 5 minute granularity, $\ell_{min}$ would equal 3. Focusing specifically on the central diagonal, SimultaneousContact$_{Total}$ is the proportional value of the number of times the two agents are in the same place at the same time, while SimultaneousContact$_{TotalRaw}$ is the raw tally of the same. SimultaneousContact$_{Sustained}$ sums the amount of sustained direct contact. In all cases, the formula is identical to its non-sustained counterpart, but constrained to the central diagonal.

In the early stages of a disease outbreak or pandemic, contact tracing using SpaRQ will have to rely on the central diagonal metrics. A risk profile can be generated simply by finding the sum of on-central diagonal recurrence; the more simultaneous contact a person has had with a contagious individual, the higher their risk assessment. If used during the current COVID-19 crisis, for example, recurrence totaling under 15 minutes could come with the recommendation to be extra vigilant with mask-wearing, hand-washing, contact with at-risk individuals, and body temperature monitoring. Exposure over 15 minutes could come with the recommendation to self-isolate for 5 to 14 days.[17] As more data is gathered, however, quantification from SpaRQ plots can be entered into regression models to help discern how the disease spreads. For instance, the distance of a recurrence point from the central diagonal can give insights into how long the virus remains viable in air, the total number of recurrence points versus sustained recurrence points can give insights into which type of contact – if either – is more likely to transmit the disease, and the number of different vectors/contagious agents can also give insights into infectivity or the presence of multiple strains. The weights from such regression models can then be used to further refine the risk assessment and pass on better information with less risk of false positives and negatives. When using SpaRQ, each new disease will present a challenge of honing the risk assessment. SpaRQ diagnostics can, however, be compared with subsequent outbreaks of previously-studied diseases to gauge mutations or social changes.

    While it is necessary to store data from both contagious individuals and healthy individuals to be able to identify trends in the SpaRQ plots that result in later positive diagnoses, it is not necessary to store data from healthy individuals to provide them with a risk assessment. Spatial data like GPS coordinates from a smartphone are only necessary to generate the latitude and longitude matrices in Figure 4. At that point, the exact position data can be deleted and the SpaRQ analysis can continue with only the binary matrices as a foundation. This addresses the security and data storage concerns that many technologists have suggested might keep part of a population from engaging in contact tracing.[9,11] If an application or public health website is constructed with SpaRQ as the basis for contact tracing and risk assessment, users could opt in to have their data saved in order to assist researchers in learning about the outbreak with the default being instant deletion of location data. Furthermore, users who opt in could be given the option

to delete their information at any time. This transparency might lead to a feeling of goodwill from the public and lead to increased use of self-guided contact tracing. The resulting anonymized SpaRQ plots, while not capable of indicating if a person eventually fell ill, could still be used to estimate infectivity by comparing the number of simultaneous and lagged proximities to recorded new cases. Lastly, technology security specialists are concerned that contact tracing applications can be used by malicious agents to pose false health emergencies like disease flare-ups. SpaRQ must have two datasets; one to test against – the contagious population – and one to test. A program using SpaRQ should not let users self-select if they are contagious or healthy; public health officials and/or hospital staff should have login credentials that allow spatial data to enter the contagious data pool. This will help prevent bad faith actors or inexperienced users from corrupting the contact tracing system.

**Conclusion**

SpaRQ is a data visualization, quantification, and analysis method that can provide disease contact tracing and individual risk assessment. Additionally, SpaRQ can cope with data from multiple sources as long as the raw input – such as latitude and longitude – is the same. In this way, both low- and high-tech contact tracing methods can be used by public health organizations to more fully canvas the outbreak area. Unlike other contact tracing methods, SpaRQ does not require data from healthy individuals to be stored; binary matrices are created from location data allowing identifiable data to be purged. Finally, like John Snow's cholera map, SpaRQ can generate data that can be used not only to manage an outbreak or pandemic, but learn about its nature, infectivity, and mutations from previous outbreaks. A simple version of SpaRQ and the sample data used in this manuscript are available on GitHub at

https://github.com/KJPatten/SpaRQ.

The author would like to thank Travis Heth, MS for assistance with the literature review.